\documentstyle[amssymb,twocolumn,eqsecnum,aps]{revtex}

\begin{document}
\draft
\title{Thermal correction to resistivity of 2D electron (hole) gas in
low-temperature measurements at $B=$0}
\author{M. V. Cheremisin}
\address{A.F.Ioffe Physical-Technical Institute, St.Petersburg, Russia}
\maketitle

\begin{abstract}
We calculate the zero magnetic field resistivity, taking into account the
degeneracy of the 2D electron (hole) gas and the thermal correction due to
the combined Peltier and Seebeck effects. The resistivity is found to be
universal function of temperature, expressed in units of $\frac{h}{e^{2}}%
(k_{F}l)^{-1}$.
\end{abstract}

\pacs{PACS numbers: 73.40.Qv, 71.30+h, 73.20.Fz}

Recently, a great deal of interest has been focussed on the anomalous
behaviour of a wide variety of low-density 2D electron$^{\text{\cite
{Kravchenko}-\cite{Popovic}}}$ and hole$^{\text{\cite{Coleridge}-\cite
{Hanein}}}$ systems, whose resistivity unexpectedly decreases as the
temperature is lowered, exhibiting a behaviour generally associated with
metals, rather than insulators. In particular, it has been found that, below
some critical 2D electron density $n_{s}^{c}$, cooling causes an increase in
resistivity, whereas at $n_{s}>n_{s}^{c}$ the resistivity decreases.
Although numerous theories have been put forward to account for this effect,
the origin of this metallic behaviour is still the subject of a heated
debate.

In the present paper, we report on a study of low-temperature transport in
2D electron gas at zero magnetic field, taking into account both the
electron degeneracy and the Peltier-effect-induced correction to resistivity.%
$^{\text{\cite{Kirby},\cite{Cheremisin}}}$ It is well known that ohmic
measurements are carried out at low current density in order to prevent
heating. Usually, only the Joule heat is considered to be important. In
contrast to the Joule heat, the Peltier and Thomson effects are linear in
current. As shown in \cite{Kirby},\cite{Cheremisin}, the Peltier effect
influences ohmic measurements and results in a correction to a measured
resistance. When current is flowing, one of the sample contacts is heated,
and the other cooled, because of the Peltier effect. The established
temperature gradient is proportional to the current. The Thomson heat is
then proportional to squared current and can therefore be neglected. Then,
the voltage drop across the circuit includes the thermoelectromotive force
induced by the Peltier effect, which is linear in current. Finally, there
exists a thermal correction $\Delta \rho $, to the ohmic resistivity, $\rho $%
, of the sample. As was demonstrated in \cite{Cheremisin}, for degenerate
electrons, $\Delta \rho /\rho \approx (kT/\mu )^{2}$ , where $\mu $ is the
Fermi energy. Hence, the above correction may be comparable with the ohmic
resistance of a sample when $kT\sim \mu ${\bf . }We further discuss the
features of thermal correction within 2D electron-density-modulated
low-temperature ohmic measurements.$^{\text{\cite{Kravchenko}-\cite{Hanein}}%
} $

Let us consider for clarity a 2DEG sample and dc current flowing in it. The
2DEG structure (MOS, quantum well, etc.) is arbitrary, electrons are assumed
to occupy the first quantum-well subband with isotropic energy spectrum $%
\varepsilon (k)=\frac{\hbar ^{2}{\bf k}^{2}}{2m}$. Here, $m$ is the electron
effective mass, ${\bf p}$=$\hbar {\bf k}$ is the electron quasi-momentum,
and ${\bf k}$ is the wave vector. The sample is connected (see Fig.\ref
{Fig.1}, insert) by means of two identical leads to the current source. Both
contacts are assumed to be ohmic. The voltage is measured between the open
ends (''c'' and ''d'') kept at the temperature of the external thermal
reservoir. The sample is placed in a sample chamber (not shown) with mean
temperature $T_{0}$.

According to our basic assumption, the contacts ( ``a'' and ``b'') may have
different respective temperatures $T_{a}$ and $T_{b}$. With the temperature
gradient term included, the current density ${\bf j}$ and the energy flux
density ${\bf q}$ are given by 
\begin{eqnarray}
{\bf j} &=&\sigma ({\bf E-}\alpha {\bf \nabla }T),\qquad  \eqnum{1}
\label{transport} \\
{\bf q} &=&(\alpha T-\zeta /e){\bf j}-\varkappa {\bf \nabla }T,  \nonumber
\end{eqnarray}
Here, ${\bf E}=\nabla \zeta /e$ is the electric field, and $\zeta =\mu
-e\varphi $ is the electrochemical potential. Then, $\sigma =Ne^{2}\tau /m$
is the conductivity, $N$ is the 2D electron concentration, $\tau $ is the
momentum relaxation time, $\varkappa $ is the thermal conductivity, and $%
\alpha $ is the 2DEG thermopower.

It is well known that the Peltier heat is generated by current flowing
across the interface between two different conductors. At the contact (
``a'' in Fig.\ref{Fig.1}, insert ), the temperature $T_{a}$, electrochemical
potential $\zeta $, normal components of the current $I=jd$, and energy flux 
$qd$ are continuous. Here, $d$ is the sample width. Then, there exists a
difference of thermopowers $\Delta \alpha =\alpha _{me}-\alpha $, where $%
\alpha _{me}$ is the thermopower of the metal lead. For $\Delta \alpha >0$
and the current direction depicted in Fig.\ref{Fig.1}, contact ``a'' is
heated, and contact ``b'' is cooled. Thus, the contacts are at different
temperatures, and $T_{a}-T_{b}=\Delta T>0$.

In general, one can easily solve Eq.(\ref{transport}), and then find $\Delta
T$ for an arbitrary circuit cooling. Since the electron-phonon coupling is
weak below $\sim $1K, the heat conduction from 2DEG to mixing chamber could
predominately occur through the ohmic contacts of the sample and the leads
connected to them. However, the experimental observations in Ref.\cite
{Mittal} demonstrate that the electron gas is, in fact, the dominant thermal
resistance in this problem. Actually, the cooling of 2D electrons with
respect to bath is provided by thermal conductivity found to follow
Wiedemann-Franz law. Accordingly, we neglect further the contact related
cooling of 2D electron gas. Then, for actual $I\rightarrow 0$ case we will
omit the Joule heating. Let us suppose for a moment that the cooling
conditions are adiabatic, with the 2D electron gas thermally insulated from
the environment. These assumptions will be later justified for an actual
2DEG system. We emphasize that under the above conditions, the sample is not
heated. Indeed, at small currents, $T_{a}\approx T_{b}\approx T_{0}$. Hence,
the amount of the Peltier heat, $Q_{a}=I\Delta \alpha T_{0}$, evolved at
contact ``a'' and that absorbed at contact ``b'' are equal. If it is
recalled that the energy flux is continuous at each contact, the difference
of the contact temperatures is given by $\Delta T=I\Delta \alpha
T_{0}l_{0}/\varkappa d$, where $l_{0}{\cal \ }$is the sample length. As
expected, $\Delta T$ is linear in current.

As shown in Ref.\cite{Cheremisin}, standard ohmic measurements always result
in a thermal correction to the resistance measured. Using Eq.(\ref{transport}%
), we find for the voltage drop $U$ between ends ``c'' and ``d'' 
\begin{equation}
U=RI+\int\limits_{c}^{d}\alpha dT,  \eqnum{2}  \label{temf}
\end{equation}
where $R$ is the total ohmic resistance of the circuit. The second term in
Eq.(\ref{temf}) coincides with the expression for the conventional
thermoelectromotive force, $\varepsilon _{T}$, under zero current
conditions. Let us assume that the temperature gradient is small. In this
case $\sigma $, $\alpha $, and $\varkappa $ can be considered constant. The
thermoelectromotive force is then given by $\varepsilon _{T}=\Delta \alpha
\Delta T$. Since $\Delta T\sim I$, there always exists a thermal correction
to the ohmic resistance $\Delta R=\varepsilon _{T}/I$. Finally, the total
resistivity of the 2DEG-sample is given by$^{\cite{Cheremisin}}$ 
\begin{equation}
\rho ^{tot}=\rho \left( 1+\frac{\alpha ^{2}}{L}\right) ,  \eqnum{3}
\label{resistivity}
\end{equation}
where $\rho =1/\sigma $ is the ohmic resistivity of the sample. In Eq.(\ref
{resistivity}) we take into account that for the actual case of metal leads $%
\Delta \alpha \simeq -\alpha $. Then, for the low-temperature case in
question we omit the phonon-related contribution to thermal conductivity.
Therefore, $\varkappa =LT\sigma ,$ where $L=\frac{\pi ^{2}k^{2}}{3e^{2}}$ is
the Lorentz number. It is noteworthy that the above result is valid for a 2D
hole gas as well.$^{\text{\cite{Cheremisin}}}$

Using the conventional Gibbs statistics$^{\cite{Landau5}}$ and the energy
spectrum specified above, the 2DEG concentration $N=-%
{\partial \Omega  \overwithdelims() \partial \mu }%
_{T}$ yields 
\begin{equation}
N=N_{0}\xi \digamma _{0}(1/\xi )  \eqnum{4}  \label{concentration}
\end{equation}
where $\Omega =-kT\sum\limits_{k}\ln (1+\exp (\frac{\mu -\epsilon _{{}}}{kT}%
))$ is the thermodynamic potential of the 2D electron gas, and $\xi =kT/\mu $
is dimensionless temperature. Then, $N_{0}=\frac{m\mu }{\pi \hbar ^{2}}$ is
the density of strongly degenerate 2DEG, and $\digamma
_{n}(z)=\int\limits_{0}^{\infty }x^{n}[1+\exp (x-z)]^{-1}dx$ is the Fermi
integral. In Fig.\ref{Fig.1}, we plot the temperature dependence of the
dimensionless concentration $n=N/N_{0}$ given by Eq.(\ref{concentration}).
In the classical Maxwell-Boltzman limit ($\xi <0,\left| \xi \right| \ll 1$),
the 2D electron density is thermally activated, and, therefore, $n=\left|
\xi \right| \exp (-1/\left| \xi \right| )$. In the case of strongly
degenerate electrons ($\xi \ll 1$), we obtain $n=1+\xi \exp (-1/\xi )$.
Then, at elevated temperatures $\xi \gg 1$, the dependence of the 2DEG
concentration $n=1/2+\xi \ln 2$ becomes linear in temperature. It is
noteworthy that the 2D electron concentration $N$ may exceed the
zero-temperature value $N_{0}$.

We emphasize that the 2D electron density is a monotonic function of
temperature (see Fig.\ref{Fig.1}). Therefore, one might expect that the
ohmic resistivity $\rho (T)\sim 1/N$ decreases with increasing temperature
at constant carrier mobility. We now demonstrate that the total resistivity
specified by Eq.(\ref{resistivity}) can, nevertheless, increase in a certain
temperature range owing to the Peltier effect-related correction.

\smallskip Following the conventional Boltzman equation formalism, the
explicit formulae for the 2DEG thermopower (for the 3D case, see Pisarenko,
1940) can be written as follows 
\begin{equation}
\alpha =-\frac{k}{e}\left[ \frac{2\digamma _{1}(1/\xi )}{\digamma _{0}(1/\xi
)}-\frac{1}{\xi }\right]  \eqnum{5}  \label{alfa}
\end{equation}
Here, we assume, for simplicity, that the electron scattering is
characterized by energy-independent momentum relaxation time. For strongly
degenerated 2DEG ($0<\xi \ll 1$), we obtain the temperature dependence of
the thermopower (Fig.\ref{Fig.1},b) as $\alpha =-\frac{k}{e}[\pi ^{2}\xi
/3-(1+3\xi )\exp (-1/\xi )]$. At elevated temperatures ($\xi >1$) the
thermopower first grows with temperature, and then approaches an universal
value $\alpha _{s}=-\frac{k}{e}\frac{\pi ^{2}}{6\ln 2}$. In the classical
Maxwell-Boltzman limit ($\xi <0,\left| \xi \right| \ll 1$) the thermopower
is given by the conventional formulae $\alpha =$ $-\frac{k}{e}(2-1/\xi )$.
Its worth noting that in the two last cases the thermopower is of the order
of $k/e$. Accordingly, the thermal correction to resistivity $\Delta \rho
=\rho \alpha ^{2}/L$ may be of the order of the 2DEG ohmic resistivity.

In Fig.\ref{Fig.2}, we plot the temperature dependence of the 2DEG
resistivity given by Eq.(\ref{resistivity}) at different Fermi temperatures $%
T_{F}=\mu /k$. At fixed temperature, the resistivity decreases with
increasing 2DEG degeneracy. Then, for a fixed Fermi energy (e.g., $T_{F}=0.5$
in Fig.\ref{Fig.2} ) the T-dependence of the resistivity exhibits metallic
behaviour at $T<T_{F}$, and then becomes insulating ( i.e.$\frac{d\rho }{dT}%
<0$ ) at $T>T_{F}$. Within the low-temperature metallic region $\xi \ll 1$,
the 2DEG resistivity can be approximated (see dashed line in Fig.\ref{Fig.2}
) with $\rho ^{tot}=\rho _{0}(1+\pi ^{2}\xi ^{2}/3)$, where $\rho _{0}{\bf =}%
\frac{h}{e^{2}}(k_{F}l)^{-1}${\bf \ }is the resistivity at $T\rightarrow 0$, 
$k_{F}=\sqrt{2m\mu }/\hbar $ is the Fermi vector, and $l=$ $\hbar k_{F}\tau
/m$ is the mean free path. Then, for the high-temperature ($\xi >1$)
insulating region we obtain the asymptote $\rho ^{tot}=\rho _{0}\frac{%
1+\alpha _{s}^{2}/L}{\xi \ln 2+1/2}$, shown in Fig.\ref{Fig.2} by dotted
line. This result is confirmed by recent experiments$^{\text{\cite{Hamilton}-%
\cite{Lewalle}}}$ shown that for temperatures well below the Fermi
temperature the metallic region data obey a scaling law where the disordered
parameter $k_{F}l$ appears explicitly. These experimental observations$^{%
\text{\cite{Hamilton}}}$ therefore rule out interactions, the shape of the
potential well spin-orbit effects as possible origins of the metallic
behaviour mechanism. Then, according to Ref.\cite{Brunthaler} the
''metallic'' state can not be associated with e-e induced quantum
interference effects. We argue that the semi-classical mechanism discussed
above may be responsible for observed T-behaviour of 2D resistivity. {\bf \ }

Let us analyze in more detail the cooling conditions, which are known to
influence the thermal correction to resistivity.$^{\text{\cite{Kirby},\cite
{Cheremisin}}}$ It will be recalled that in the case of adiabatic cooling
the electron temperature differs from the bath temperature $T_{0}$. We now
consider the opposite situation of electron cooling due to, for example,
finite strength of electron-phonon coupling. Let the phonons are thermalized
with respect to bath temperature $T_{0}${\bf .} Following Ref.\cite{Sivan},
below $\sim $0.6K in Si MOSFET's the electron-to-phonon thermal exchange is
given by $a(T^{3}-T_{0}^{3})$, where $a=$2.2 $\times $10$^{-8}$W/K$^{3}$cm$%
^{2}$. When $T-T_{0}\ll T_{0}$, the thermal correction to the resistivity $%
\Delta \rho $ is suppressed$^{\text{\cite{Cheremisin}}}$ by the factor $%
\beta =\frac{\tanh \lambda }{\lambda }$, where $\lambda =l_{0}T_{0}\sqrt{%
3a/4\varkappa }$ is a dimensionless parameter. Actually, $\lambda $ is the
ratio of outgoing and internal heat fluxes associated with phonon related
thermal leakage and electron heat diffusion, respectively.

When $\lambda \ll 1$, the local cooling due to phonons can be neglected and
the adiabatic approach is well justified. In the opposite case of intensive
cooling ($\lambda \gg 1$), the difference of the contact temperatures $%
\Delta T$ becomes smaller, and, therefore, the thermal correction to
resistivity $\Delta \rho $ vanishes. For $l_{0}=1$mm, $T_{0}=50$mK, $\sigma
=2e^{2}/h=8\times 10^{-5}$Ohm$^{-1}$( typical critical region conductance)
we obtain $\lambda =1.2$, and, therefore, $\beta =0.8$. It worthwhile to
notice that Peltier correction to resistivity becomes greater at ultra-low
temperatures for short samples, since $\lambda \sim l_{0}T_{0}$. In real
experiments both the electron-phonon coupling and the sample-to-bath thermal
exchange may be important.

We emphasize that both dc and ac ohmic measurements lead to a thermal
correction. The correction is, however, strongly damped at high frequencies
because of the thermal inertial effects. As demonstrated in Ref.$\cite{Kirby}
$, the above quasi-static approach is valid below some critical frequency $%
f_{cr}=\chi /l_{0}^{2}$. For example, for degenerate electrons the thermal
diffusion coefficient $\chi $ is of the order of the diffusion coefficient $%
D=\frac{\sigma }{e^{2}}\left( \frac{dN_{0}}{d\mu }\right) ^{-1}$. Assuming $%
\sigma =e^{2}/h$, $l_{0}=1$mm, for GaAs-based structure we obtain $\chi \sim
\hbar /m$, hence $f_{cr}=1.5$kHz. We suggest that the spectral dependence of
the 2D resistivity can be used to estimate the thermal correction.

In conclusion, low-temperature ohmic measurements of a 2D electron (hole)
gas at $B=0$ involve a thermal correction caused by the Peltier effect. The
magnitude of thermal correction depends on the 2DEG degeneracy and actual
cooling conditions. The resistivity of 2DEG with thermal correction included
is found to be universal function of temperature, expressed in units $%
h/e^{2}(k_{F}l)^{-1}$. This universal behaviour correlates with that found
in experiments.

This work was supported by RFBR, INTAS (grant YSF 2001/1-0132), and the LSF
program(Weizmann Institute). The author acknowledges the generous
hospitality of the Institut fur Theorie der Kondensierten Materie, Karlsruhe.

\smallskip 
\begin{figure}[tbp]
\caption{Themperature dependence of 2DEG concentration (a) and thermopower
(b) given by Eq.(\ref{concentration}) and Eq.(\ref{alfa}) respectively.
Asymptotes: $\left| \xi \right| \ll 1$ - dotted line, $\xi \gg 1$ - dashed
line. Inserts: the experimental setup (top); position of the Fermi level
with respect to the bottom of quantum-well subband (bottom).}
\label{Fig.1}
\end{figure}

\begin{figure}[tbp]
\caption{T-dependence of the 2DEG resistivity, given by Eq.(\ref{resistivity}%
-\ref{alfa}) for $T_{F}=2;1.75;1.5;1.25;1;0.75;0.5;$ $%
0.3;0.25;0.2;0.15;0.1;0.05;0.01$K. Asymptotes: $\xi \ll 1$ - dashed line, $%
\xi >1$ - dotted line for $T_{F}=0.5$K. Insert: density dependence of the
2DEG resistivity within the $T=$0.5-0.9K range.}
\label{Fig.2}
\end{figure}

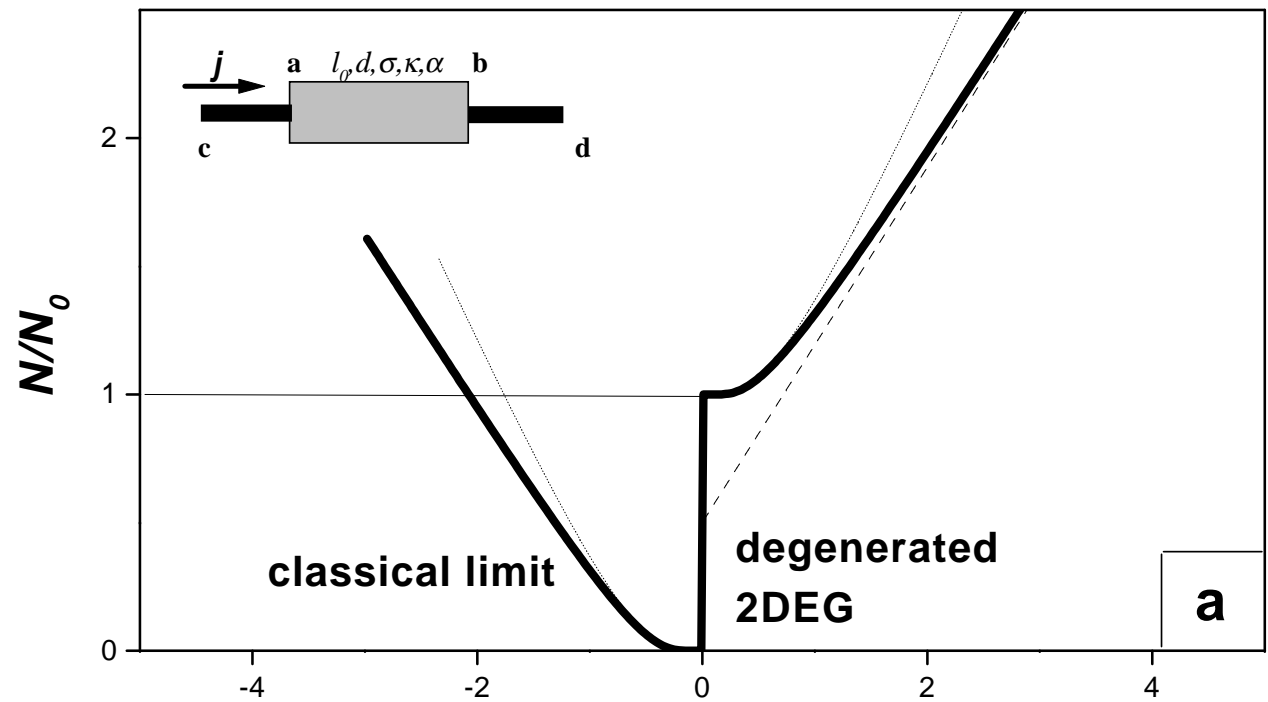
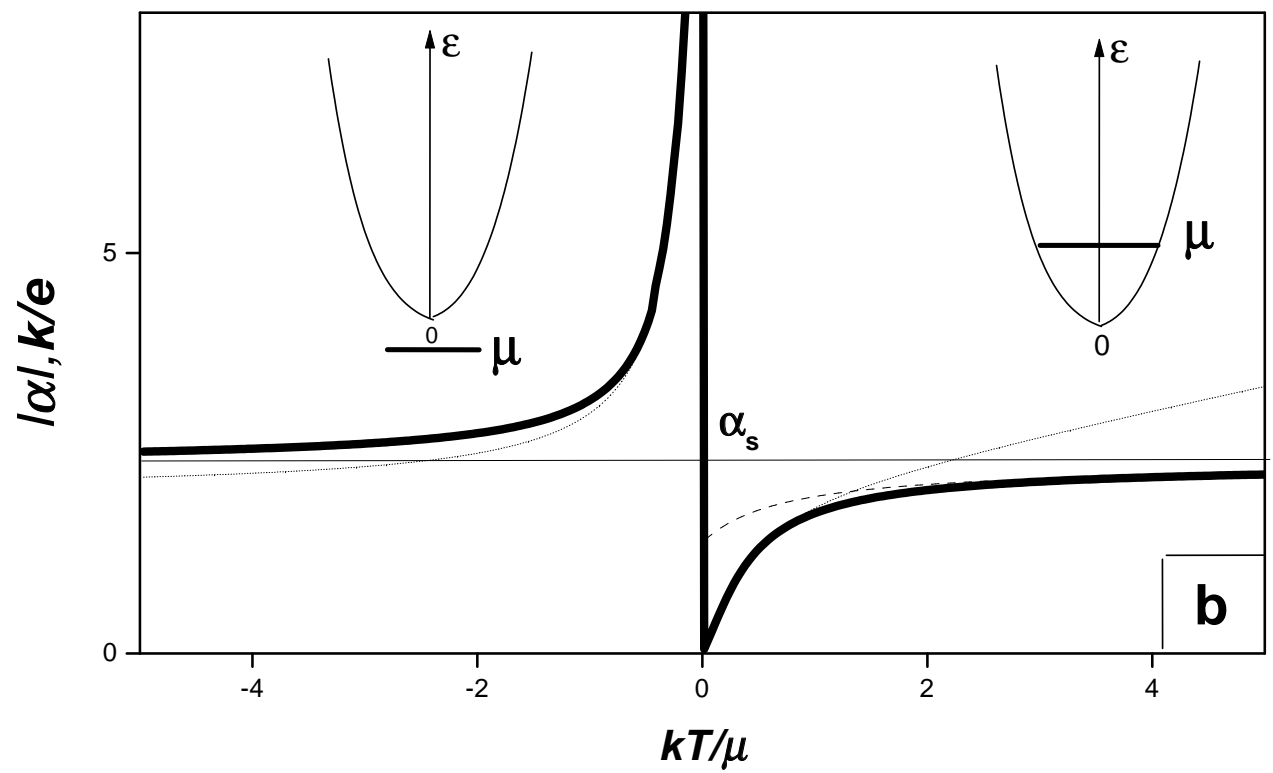
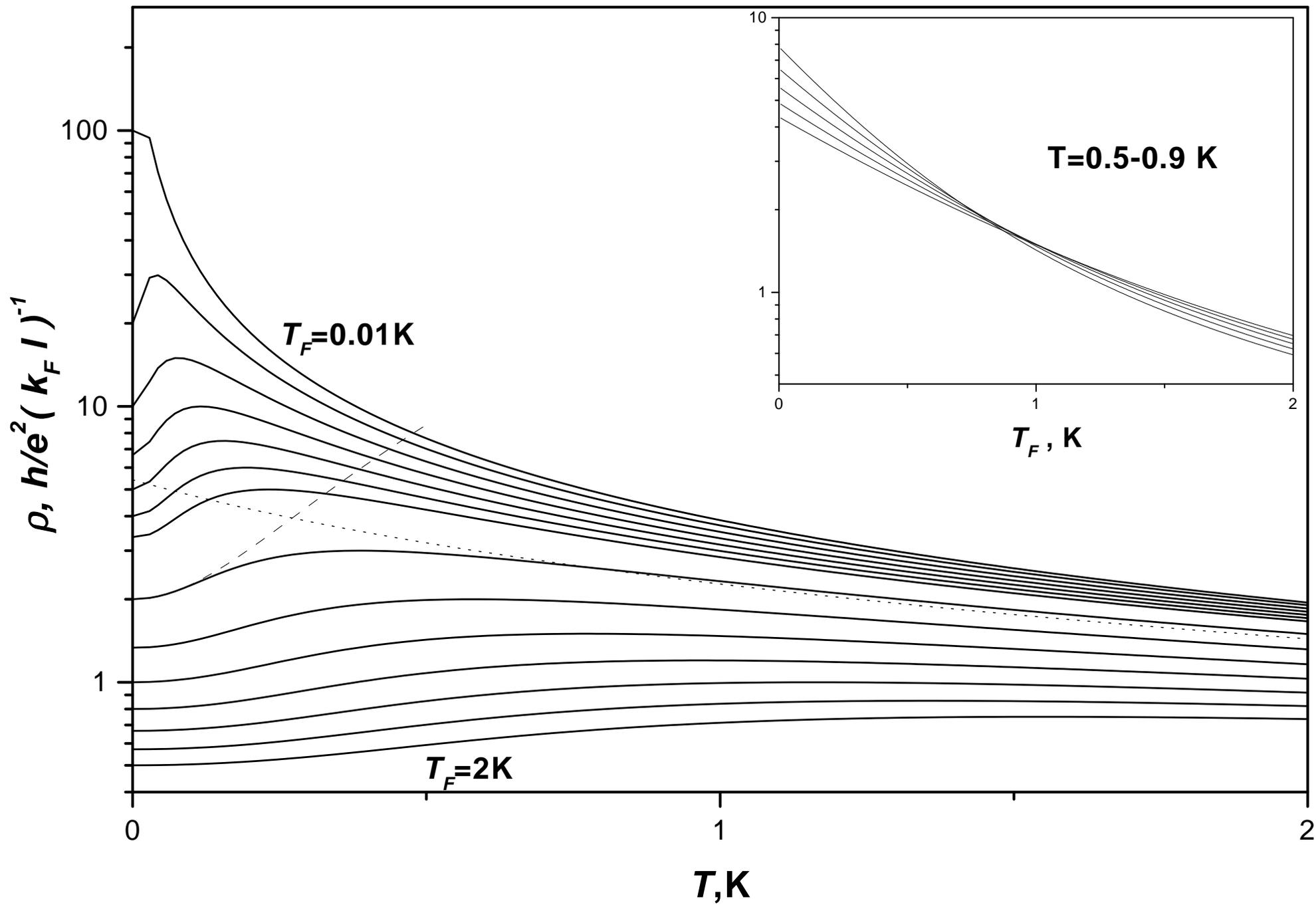
\end{document}